\input{aipcheck}

\documentclass[
    ,final            % use final for the camera ready runs
  ]
  {aipproc}

\layoutstyle{6x9}

\begin{document}

\title{Gluon Polarization Measurements from Correlated Probes}

\classification{14.20.Dh, 13.87.Ce, 13.88.+e, 14.70.Dj }
\keywords      {gluon polarization, dijet production, STAR, RHIC}

\author{Matthew Walker for the STAR collaboration}{
  address={Massachusetts Institute of Technology,
Laboratory for Nuclear Science,
Cambridge, MA 02139}
}

\begin{abstract}
These proceedings show the preliminary results of the dijet longitudinal double spin asymmetry $A_{LL}$ in polarized proton-proton collisions at $\sqrt{s} =  200$ GeV for the pseudorapidity range $|\eta| \le 0.8$. The integrated luminosity of 10.6 pb$^{-1}$ used in this analysis was collected during RHIC Run-9. This result is presented as a function of the dijet invariant mass in multiple pseudorapidity acceptances. The division into different pseudorapidity regions allows the kinematics of the hard interaction to be constrained and information about the shape of $\Delta g(x)$ to be extracted for the first time. Comparisons are made with expectations from various different theoretical scenarios of polarized parton distributions of the proton.
\end{abstract}

\maketitle

%%%%%%%%%%%%%%%%%%%%%%%%%%%%%%%%%%%%%%%%%%%%
%% MAINMATTER
%%%%%%%%%%%%%%%%%%%%%%%%%%%%%%%%%%%%%%%%%%%%

\section{Introduction}
The proton is a composite particle made up of quarks and gluons whose interactions are described by quantum chromodynamics (QCD). The spin structure of the proton remains an area of intense study by both experimentalists and theorists. It can be described in a simple picture as
\begin{equation}
\frac{1}{2} = \frac{1}{2} \Delta \Sigma + L_q + \Delta G + L_g
\end{equation}
where $\Delta \Sigma$ represents the quark spin contribution, $\Delta G$ represents the gluon spin contribution, and $L_{q(g)}$ represents the quark (gluon) orbital angular momentum. This relation can actually be constructed in QCD in the infinite momentum frame and the light cone gauge \cite{Jaffe1990509}.

Polarized deep inelastic scattering (DIS) experiments indicate that quarks only carry approximately 30\% of the proton spin, which led to interest in measuring the contributions from the remaining components. Polarized proton-proton collisions at the Relativistic Heavy Ion Collider (RHIC), the world's only polarized proton collider, permit measurements that allow an extraction of the polarized gluon distribution function $\Delta g(x)$. This function describes the contribution of the gluon polarization to the proton spin and is poorly constrained by DIS data. Measurements of the longitudinal double-spin asymmetry ($A_{LL}$) can be used to improve understanding of $\Delta g(x)$. Previous measurements of the dijet cross section at RHIC find good agreement with pQCD predictions \cite{1742-6596-295-1-012068}.

$A_{LL}$ is defined as the ratio of the polarized to unpolarized cross section of a process and can be defined at RHIC according to Equation \ref{eqn:all}, where $N^{++(+-)}$ is the yield when the colliding proton helicities are aligned (anti-aligned), $R$ is the relative luminosity of the two configurations, and $P_{B(Y)}$ is the polarization of the blue (yellow) RHIC beam.
\begin{equation}
\label{eqn:all}
A_{LL} = \frac{d\Delta\sigma}{d\sigma} = \frac{1}{P_BP_Y}\frac{N^{++}-RN^{+-}}{N^{++}+RN^{+-}}
\end{equation}

Approximately 10.6 pb$^{-1}$ of RHIC data collected in 2009 at the Solenoidal Tracker at RHIC (STAR) were used in this analysis. The average beam polarization during this run was 58\%, which is determined in RHIC using a combination of proton-carbon Coulomb-nuclear interference (CNI) polarimeters \cite{Spinka:2003p179} and a hydrogen gas-jet polarimeter  \cite{Okada:2008p764}. The STAR detector \cite{Ackermann2003624} is able to reconstruct jets using the Time Projection Chamber (TPC) and the Barrel Electromagnetic Calorimeter (BEMC). The TPC provides charged particle track reconstruction in a 0.5 T magnetic field over a pseudorapidity range of $|\eta| < 1.3$ and over full azimuth. The BEMC provides electromagnetic energy reconstruction over a pseudorapidity range of $|\eta| < 1$ and over full azimuth. The Beam-Beam Counters (BBCs) and Zero-Degree Calorimeters (ZDC) provide luminosity monitoring capabilities.

\section{Analysis and Results}
A midpoint-cone algorithm \cite{cdfmidpointcone} was used to reconstruct jets with cone radius of 0.7 in $\eta-\phi$ coordinates. The split-merge fraction was set to 0.5, and the seed threshold was set to 0.5 GeV in transverse energy for either tracks or towers. Events with more than one jet were eligible for inclusion in the dijet analysis.

The dijet asymmetry was measured for dijets with three different pseudorapidity topologies: two jets have the same sign pseudorapidity, two jets have opposite sign pseudorapidity and the inclusive measurement. The combination of these acceptances and the dijet invariant mass bins provides access to the initial Bjorken-$x$ kinematics of the colliding protons according to the relations:
\begin{eqnarray}
M&=&\sqrt{x_1 x_2 s} \nonumber \\ \nonumber \\
\frac{\eta_3 + \eta_4}{2} &=& \frac{1}{2}\ln(\frac{x_1}{x_2}),
\end{eqnarray}
where $x_1$ and $x_2$ are the initial parton Bjorken-$x$ values, $\sqrt{s}$ is the center-of-mass energy of the collision, $M$ is the dijet invariant mass, and $\eta_3$ and $\eta_4$ are the pseudorapidities of the measured jets.

Events containing jets that satisfy the acceptance conditions must be back-to-back ($\Delta \phi > 2$) and fulfill a theoretically motivated \cite{asymmcut} asymmetric $p_T$ cut (max($p_T$) > 10 GeV/c and min($p_T$) > 7 GeV/c). One of the jets must point at the region of the BEMC that triggered the event.

The dijet yields are sorted according to the helicity configuration of the colliding protons for the relevant bunch crossing. The spin-sorted yields are unfolded according to a matrix unfolding technique \cite{bayesunfolding} to account for bin shifting that results from finite detector resolution. The unfolding matrix was calculated using simulations done on a cloud computing facility at Clemson University \cite{clemson}. Events were generated using {\sc Pythia} version 6.4.23 \cite{pythia} with tune 329 (Pro-pt0) \cite{Skands:2010p2204} and the detector response simulated using {\sc Geant} \cite{geant}. Event filtering was used to substantially decrease the amount of time needed to generate the simulation. Comparisons between data and simulation kinematic distributions showed good agreement.

\begin{figure}
  \includegraphics[scale=0.41]{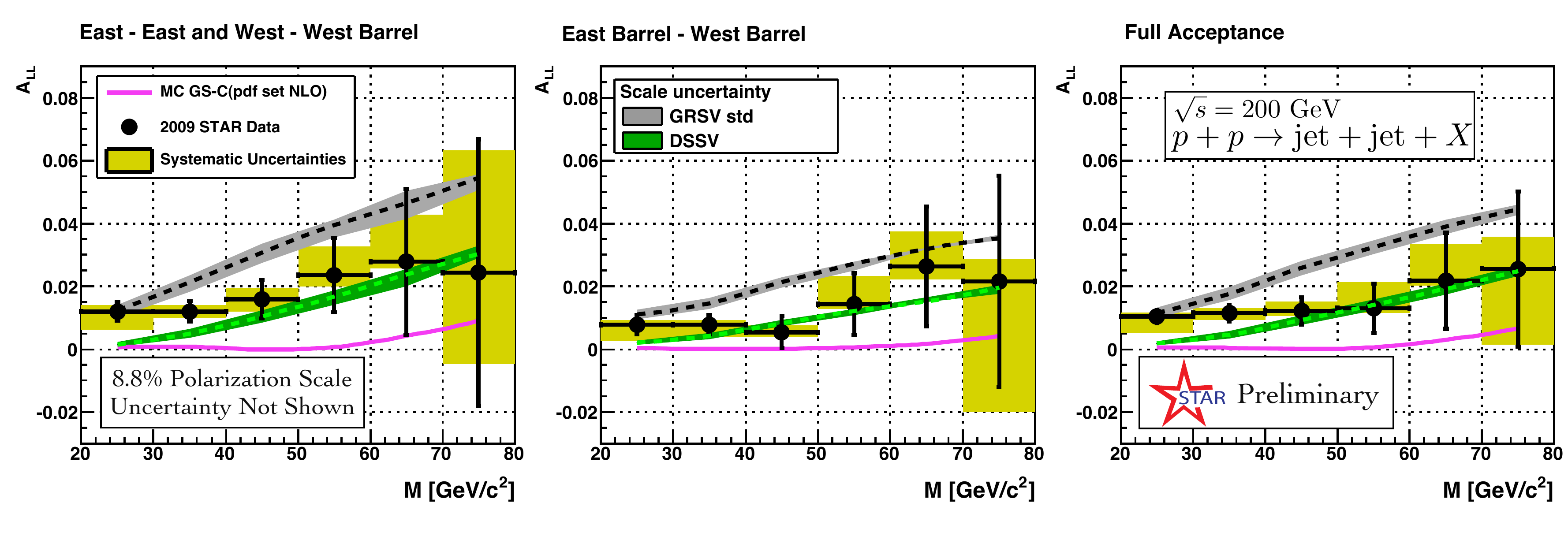}
  \caption{\label{fig:result}The 2009 dijet $A_{LL}$ measured at STAR has been measured in three pseudorapidity acceptances, which allows the result to be used to constrain the kinematics of the interacting partons.}
\end{figure}

The asymmetries for the three different acceptances can be seen in Figure \ref{fig:result}. The black points with error bars are the data with statistical uncertainties, while the systematic uncertainties are shown as yellow bands. The dominant systematics are the uncertainty in the trigger efficiencies which arises from the different sub-process mixtures in the different theory scenarios and the uncertainty in the jet energy scale. The results are compared with the calculations of GRSV (std) \cite{Gluck:2000dy}, DSSV \cite{deFlorian:2008mr}, and GS (C) \cite{Gehrmann:1995ag}.

\begin{figure}
  \includegraphics[scale=0.45]{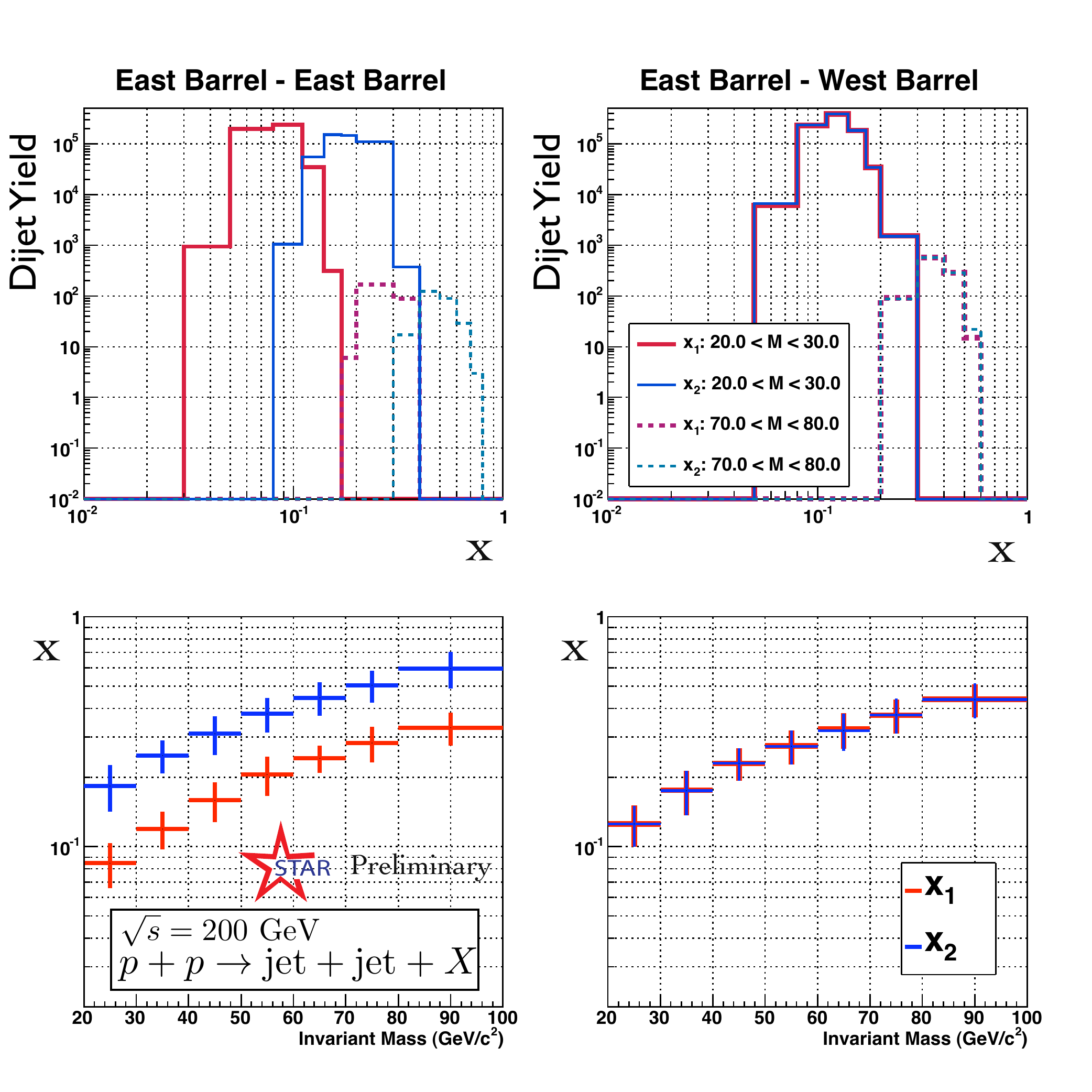}
  \caption{\label{fig:kinematics}The kinematic coverage of the two exclusive acceptance bins will allow shape constraints to be placed on $\Delta g(x)$.}
\end{figure}

The kinematics represented by the two exclusive acceptances are shown in Figure \ref{fig:kinematics}. The left panels show the kinematics for the topology with the jets having the same sign pseudorapidity (represented by the East or West side of the detector) and the right panels show them for the jets having opposite sign. The top panels show the distributions of $x_{1(2)}$ in red (blue) for a low (high) invariant mass bin in the solid (dashed) lines. The bottom panels show the mean and RMS of the $x_{1(2)}$ distributions in red (blue).

\section{Conclusions}
The dijet longitudinal double-spin asymmetry $A_{LL}$ has been measured in polarized proton-proton collisions at $\sqrt{s}$ = 200 GeV by the STAR experiment using data from the 2009 RHIC running period. The measurement of the dijet $A_{LL}$ in multiple pseudorapidity acceptances allows constraints to be made on the shape of $\Delta g(x)$ for the first time. These constraints will improve the uncertainties on extrapolations of $\Delta g(x)$ to lower $x$ kinematic regions. This result, along with the STAR inclusive jet result from 2009 \cite{piberoDIS2011}, represents the first measurement of a non-zero $A_{LL}$ at mid-rapidity.

%%%%%%%%%%%%%%%%%%%%%%%%%%%%%%%%%%%%%%%%%%%%
%% Sample figure:
%%
%% The option [height=...] scales the picture to the given height,
%% without it it would be printed at its nominal size
%%%%%%%%%%%%%%%%%%%%%%%%%%%%%%%%%%%%%%%%%%%%

%\begin{figure}
%  \includegraphics[scale=0.5]{2009dataSimuComp}
%  \caption{\label{fig:dataSimuComp}}
%\end{figure}

\bibliographystyle{aipproc}   % if natbib is available

\bibliography{DIS2011_dijets}

\IfFileExists{\jobname.bbl}{}
 {\typeout{}
  \typeout{******************************************}
  \typeout{** Please run "bibtex \jobname" to optain}
  \typeout{** the bibliography and then re-run LaTeX}
  \typeout{** twice to fix the references!}
  \typeout{******************************************}
  \typeout{}
 }

\end{document}